\documentclass[10pt,journal,compsoc]{IEEEtran}

\ifCLASSOPTIONcompsoc
  \usepackage[nocompress]{cite}
\else
  \usepackage{cite}
\fi
\ifCLASSINFOpdf
\else
\fi
\usepackage{kotex}
\usepackage{verbatim}
\usepackage{multirow}
\usepackage{amsmath}
\usepackage{microtype}
\usepackage{graphicx}
\usepackage{subfigure}
\usepackage{booktabs} 
\usepackage[
  separate-uncertainty = true,
  multi-part-units = repeat
]{siunitx}
\usepackage{tabularx}

\usepackage{algorithm}
\usepackage{algpseudocode}

\begin{document}
%
\title{CIMS: Correction-Interpolation Method\\ for Smoke Simulation}
%
%
%
%

\author{Yunjee~Lee,
        Dohae~Lee,
        Young~Jin~Oh,
        and~In-Kwon~Lee 

        
\IEEEcompsocitemizethanks{\IEEEcompsocthanksitem Yunjee Lee, Dohae Lee and In-Kwon Lee are with the Department of Computer Science, Yonsei University, Seoul, Korea. E-mail: yoong15, dlehgo1414, iklee@yonsei.ac.kr. The corresponding author is In-Kwon Lee.
\IEEEcompsocthanksitem Young Jin Oh is with ICT Advanced Robotics Lab, LG Electronics. E-mail: skrcjstk@gmail.com.}%
\thanks{}}

%
%

\markboth{}{}
\IEEEtitleabstractindextext{%
\begin{abstract}
In this paper, we propose CIMS: a novel correction-interpolation method for smoke simulation. The basis of our method is to first generate a low frame rate smoke simulation, then increase the frame rate using temporal interpolation. However, low frame rate smoke simulations are inaccurate as they require increasing the time-step. A simulation with a larger time-step produces results different from that of the original simulation with a small time-step.	
Therefore, the proposed method corrects the large time-step simulation results closer to the corresponding small time-step simulation results using a U-Net-based DNN model.
To obtain more precise results, we applied modeling concepts used in the image domain, such as optical flow and perceptual loss.
By correcting the large time-step simulation results and interpolating between them, the proposed method can efficiently and accurately generate high frame rate smoke simulations.
We conduct qualitative and quantitative analyses to confirm the effectiveness of the proposed model.
Our analyses show that our method reduces the mean squared error of large time-step simulation results by more than 80\%  on average.
Our method also produces results closer to the ground truth than the previous DNN-based methods; it is on average 2.04 times more accurate than previous works.
In addition, the computation time of the proposed correction method barely affects the overall computation time.
\end{abstract}

}

\maketitle

\IEEEdisplaynontitleabstractindextext

%
\IEEEpeerreviewmaketitle

\ifCLASSOPTIONcompsoc
\IEEEraisesectionheading{\section{Introduction}\label{sec:introduction}}
\else
\section{Introduction}
\label{sec:introduction}
\fi

%
%
%
%
\IEEEPARstart{S}{moke} simulation, along with water and fire simulation, is conducted based on fluid simulation and plays an important role in natural scenes created with computer graphics.
However, conducting smoke simulations requires extremely high overheads, since it involves solving complex formulas based on Navier-Stokes equations.
Reducing the enormous computational cost of precise simulations has been a challenging problem.

Various studies have made efforts to improve traditional physics-based methods \cite{4015404,  solenthaler2011two,golas2012large,7127055,ando2015dimension,yan2016multiphase, 7845705, sato2021stream}.
Recently, Deep Neural Network (DNN)-based methods have been actively proposed to efficiently compute fluid simulations \cite{tompson2017accelerating, xie2018tempogan, werhahn2019multi, bai2020dynamic, sanchez2020learning, roy2021neural}.
Especially, directly generating the simulation results with a DNN model dramatically improved the speed of simulations.
DNN models that directly predict velocity fields \cite{kim2019deep, Chu2021Learning} omit calculating complex physics equations. These methods, however, require forward advection of densities at every step. That is, repeating DNN inference and simulations is needed for every frame generation.
On the other hand, models predicting the temporal evolution of fluid using latent space representations, which do not require additional forward advection to obtain density fields, were also proposed \cite{wiewel2019latent, wiewel2020latent}.
These methods completely removed the use of a physics-based solver.
However, they still suffer from the same limitations as physics-based methods. 
They require the sequential generation of frames because the information of the previous frame is necessary to make the subsequent frame.
The generation of the frames cannot be parallelized due to this dependency between each step.

\begin{figure}[t]
  \centering
  \includegraphics[width=1.0\linewidth]{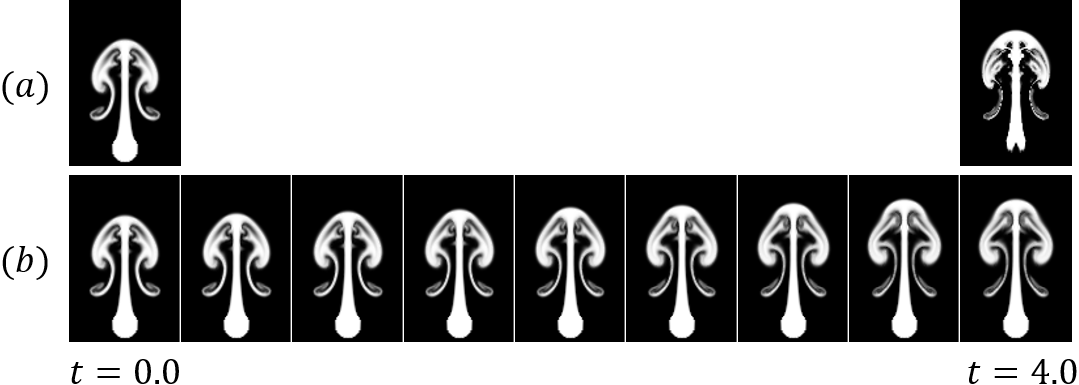}
  \caption{\label{fig:diff_smallbig}
           Sample results from the large time-step and small time-step simulations. Despite being computed from the same initial state, the result of \textnormal{(a)} large time-step simulations and \textnormal{(b)} corresponding  small time-step simulations are different.}
\end{figure}

Oh and Lee \cite{oyj} proposed the Two-Step temporal Interpolation method (TSI) to solve this problem.
TSI first generates a low frame rate smoke simulation, then increases the frame rate using temporal interpolation.
The calculation using a physics-based solver is required for only a few frames in the low frame rate simulation, which are then used as inputs to the interpolation.
Intermediate frames between the input frames are quickly generated through DNN inference.
In addition, these intermediate frames are independent of each other, making it possible for the inference to be fully parallelized.
Thus, frames can be produced quickly despite the use of the solver.
Nevertheless, TSI is not without limitations.
It assumed that the simulation results computed with different time-step sizes are the same.
However, simulations computed with small time-step are more accurate than large time-step simulations (Figure \ref{fig:diff_smallbig}).
Therefore, the method's interpolation results are different from the corresponding target small time-step simulation results. 

In this paper, we propose CIMS: a novel correction-interpolation method for smoke simulation.
The proposed method efficiently conducts smoke simulations by repeating the following steps.
First, a physics-based solver computes a single step of a large time-step simulation.
Then the U-Net\cite{ronneberger2015u} structured correction network corrects the large time-step simulation result to be close to the corresponding small time-step simulation result.
In particular, for the density field correction, optical flow-based warping and direct prediction are combined to make the result clearer.
Finally, the interpolation network interpolates between the initial state and the corrected result.
By interpolating between the corrected frames, it can produce intermediate frames close to the smaller time-step simulation results. 

Through experiments, we confirm that the proposed method effectively corrects the large time-step simulation results. 
Our method reduces the mean squared error of the large time-step simulations by more than 80\% on average in most test datasets.
The method also produces closer results to the ground truth than the previous DNN-based methods.
The error of the results is on average 2.08 times lower than other works and 3.18 times more accurate than the previous naive interpolation method \cite{oyj}.
Moreover, despite adding the correction process on top of the interpolation process, our method is still, on average, three times faster than the traditional physics-based method.

\section{Related Works}
\label{sec:relatedwork}
\subsection{Fluid Simulation using Deep Neural Networks}
Efficient fluid simulation has always been a topic of interest in computer graphics.
Various studies have made efforts to improve traditional physics-based methods \cite{4015404,  solenthaler2011two,golas2012large,7127055,ando2015dimension,yan2016multiphase, 7845705, sato2021stream}.
Recently, DNN-based methods have been actively proposed  to compute fluid simulations efficiently \cite{sanchez2020learning, roy2021neural}.
Many of the complex computations required for simulation can be omitted with DNNs.
Some replace the pressure projection part of the simulation with Convolution Neural Networks (CNN) \cite{tompson2017accelerating, xiao2018adaptive}.
CFDNet \cite{obiols2020cfdnet} improved CFD \cite{ragheb1976computational} solver by accelerating the convergence of simulations using CNN.
These methods focus on reducing the time it takes for the physics-based solver to create a single frame, requiring DNN inference and simulations for each frame generation.
Another approach is to directly generate the simulation result. 
For example, a CNN-based autoencoder could be used to generate velocity fields \cite{kim2019deep}, or a GAN-based model could learn meaningful controls of simulations from density fields \cite{Chu2021Learning}.
These methods yet need a forward advection process with the predicted velocity fields to obtain density fields, which is the final visible result of the simulation.
Generative methods that predict the temporal evolution of the fluid based on latent space representation do not require any additional advection \cite{wiewel2019latent, wiewel2020latent}.
In particular, Wiewel et al.
\cite{wiewel2020latent} encodes the velocity field and density field to a latent representation.
With the two consecutive encoded latents, an LSTM-based temporal predictor predicts the latent of the next step. 
The model finally decodes the latent to the velocity field and density field of the next step.
Since the decoded density field result can be used immediately, no additional advection process is required.
In other words, simulation results can be generated only with DNN model inference without a solver.
However, these methods still have similar limitations as to when using a physics-based solver.
The information of the previous step is required to generate the result of the next step.
The generation of the frames cannot be parallelized because the future frame generation is dependent on the previous frame.

Some recent DNN-based works use interpolation methods for efficient simulation. Interpolation methods can be classified into spatial interpolation and temporal interpolation. 
Spatial interpolation applies super-resolution problem to fluid simulation; it conducts simulation at a small resolution and then upscales it to a larger resolution \cite{xie2018tempogan, werhahn2019multi, bai2020dynamic, 9082171}.
Temporal interpolation conducts simulation in large steps and then generates the intermediate frames through interpolation.
Oh et al.\cite{oyj} applied the temporal interpolation method, which is usually used in videos, to smoke simulations. This method generates high frame rate smoke frames from low frame rate ones by applying temporal interpolation to the density fields. Although the use of the physics-based solver is required, it generates high frame rate smoke frames more quickly than the previous methods because the generation of the intermediate frame can be fully parallelized.
However, this method overlooks the fact that the results produced by the semi-Lagrangian method with large time-steps are inaccurate.
Interpolating between these inaccurate smoke frames produces results different from those computed with small time-steps.

A recent work that effectively models turbulent flow \cite{10.1145/3478513.3480492} uses spatio-temporal interpolation, which is the combination of the two kinds of interpolation. However, this interpolation method does not remove the dependencies between the intermediate frames; therefore, it cannot be fully parallelized. The inaccuracy of large time-step simulations was also not considered.
To overcome the limitation of the previous works, we propose a method that corrects the large time-step simulation results before applying interpolation.

\subsection{Image Generation using Optical Flow}\label{sec:img_g_optical_flow}
Studies on generating high-quality images using DNN have been actively conducted.
Architectures based on VAE \cite{vae} and GAN \cite{gan} have shown successful results in image generation tasks \cite{gulrajani2016pixelvae, huang2018introvae, karras2019style, choi2020stargan}.
However, it is challenging to keep the resulting image from becoming blurry or forming artifacts  \cite{cai2019multi, brock2018large}.
Video generation tasks are even more difficult because the network should learn to model both appearance and motion patterns \cite{jiang2018super}.
Optical flow is often used to overcome these difficulties \cite{Liu_2017_ICCV, chen2018lip, Xu_2019_CVPR, Wu_2020_CVPR}.
SuperSloMo \cite{jiang2018super} generates intermediate frames for temporal interpolation by combining the warping results of two frames based on the optical flow learned through unsupervised learning.
Fusing pixel generation and optical flow estimation are also widely used to produce more sophisticated frames \cite{Liang_2017_ICCV, wang2018video, lu2021video}.
In particular, Vid2Vid \cite{wang2018video} produces more sophisticated frames from the weighted sum of the warped image using optical flow and the image generated by a GAN-based model.
It is also possible to generate video frames by fusing the predicted optical flow and textures  \cite{ohnishi2018hierarchical, Li_2018_ECCV}.
Another method generates video by decoding the optical flow of the encoded motion obtained from a single semantic label map \cite{pan2019video}.
Wang et al. \cite{wang2020vidsr} use the optical flow of low-resolution frames to generate high-resolution video.
Besides video generation tasks, optical flow is also used in tasks where warping through feature matching is efficient. For example, one of the facial image frontalization techniques uses bi-directional flow fields that represent pixel correspondence between profile and frontal face \cite{wei2020learning}.

Likewise, we adopt this idea to correct density fields to make more precise results in our method.
When the imprecise density field computed with large time-step is appropriately advected forward, it will become much closer to the density field of the corresponding small time-step simulation result.
Therefore, we can assume that there is a flow between the input density field and the target ground truth.
Our method utilizes this optical flow, in addition to directly predicting the density fields.
By combining them, the network can effectively model both temporal and spatial information.

\section{Methods}
\label{sec:methods}
    \begin{figure}[t]
        \centering
        \includegraphics[width=1.\linewidth]{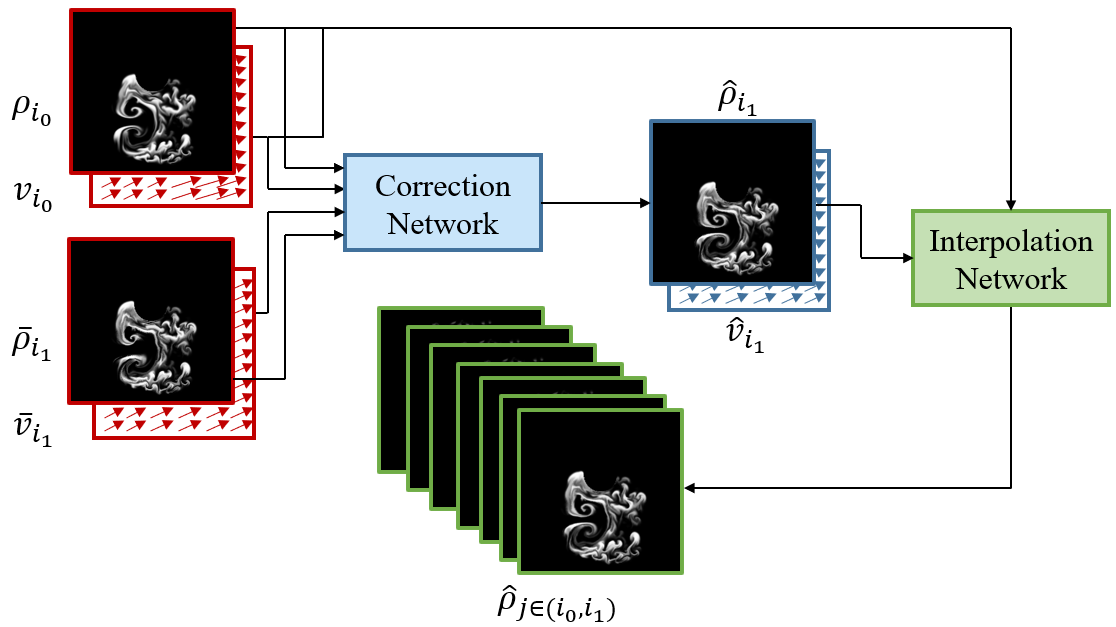}
        \caption{\label{fig:overview}
           Overview of the proposed method. $\bar{\rho}_{i_1}$ and $\bar{v}_{i_1}$ denote the density and velocity fields of the large time-step simulation result computed from given density and velocity fields ${\rho}_0$ and ${v}_0$, respectively. $\hat{\rho}_{i_1}$ and $\hat{v}_{i_1}$ denote the correction results of $\bar{\rho}_{i_1}$ and $\bar{v}_{i_1}$, respectively.}
\end{figure}

    Figure \ref{fig:overview} shows the overview of our proposed method. $i_0$ is the input frame index, and $i_1$ is the index after one large time-step from $i_0$. Given the input density field ${\rho}_{i_0}$ and velocity field ${v}_{i_0}$, we first use the physics-based solver to compute the density field $\bar{\rho}_{i_1}$ and velocity field $\bar{v}_{i_1}$ with a large time-step.
    The correction network then corrects the density and velocity fields $\bar{\rho}_{i_1}$ and $\bar{v}_{i_1}$ to $\hat{\rho}_{i_1}$ and $\hat{v}_{i_1}$, respectively. Finally, the interpolation network interpolates between ${\rho}_{i_0}$ and $\hat{\rho}_{i_1}$ to produce the intermediate frames.
    We introduce our correction network in Section \ref{subsec:correction_network} and describe the interpolation network in Section \ref{subsec:interpolation network}.
    While training the networks, we used the target small time-step simulation as the ground truth.
    
\subsection{Correction Network} \label{subsec:correction_network}
    \begin{figure}[tp]
          \centering
          \includegraphics[width=.8\linewidth]{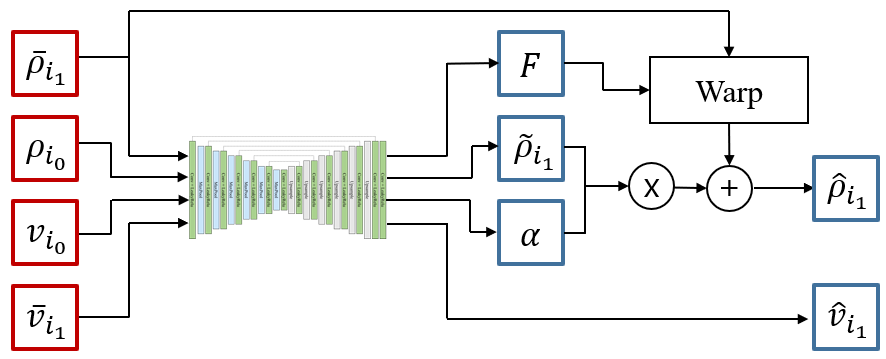}
          \caption{\label{fig:correction_network}
                   Structure overview of the correction network. Given ${\rho}_{i_0}$, ${v}_{i_0}$, $\bar{\rho}_{i_1}$, $\bar{v}_{i_1}$ as inputs, the network directly outputs the corrected velocity field $\hat{v}_{i_1}$ and outputs $F$, $\tilde{\rho}_{i_1}$, $\alpha$ to generate the corrected density field $\hat{\rho}_{i_1}$.}
        \end{figure}
    The correction network (Figure \ref{fig:correction_network}) corrects $\bar{{v}}_{i_1}$ and $\bar{\rho}_{i_1}$, the results of the large time-step simulation, to be closer to the corresponding small time-step simulation results.
    The inputs of the correction network are ${v}_{i_0}$, ${\rho}_{i_0}$, $\bar{{v}}_{i_1}$, and $\bar{\rho}_{i_1}$, and the total objective function to optimize the correction network is defined as:
    \begin{equation}\label{res_img}
                    L = L_{v} +  L_{\rho},
    \end{equation}
    where $L_{v}$ and $L_{\rho}$ denote velocity correction and density correction loss, respectively. 
    
    \subsubsection{Velocity Correction}\label{subsubsec:velocity_correction}
    The correction network directly predicts $\hat{v}_{i_1}$, which is the corrected version of the input $\bar{v}_{i_1}$.
    The loss function to optimize the velocity correction $L_v$ is defined as:
    \begin{equation}\label{res_img}
                    L_{v}=\lambda_{r_v}L_{r_{v}} + \lambda_{g}L_{g} + \lambda_{p}L_{p_{v}} + \lambda_{t_v}L_{t_v}.
    \end{equation}
    $L_{r_{v}}$, $L_{g}$, $L_{p_{v}}$, and  $L_{t_{v}}$ denote the reconstruction loss, gradient loss, perceptual loss, and temporal coherence loss of the velocity fields, respectively. $\lambda_{r_{v}}$, $\lambda_{g}$, $\lambda_{p_{v}}$, and $\lambda_{t_{v}}$ are the weights of each term.
    
    \textit{Velocity Reconstruction Loss.}
    The reconstruction loss for velocity field models how close the reconstructed velocity field is to the ground truth. The velocity reconstruction loss $L_{r_{v}}$ is defined as:
    \begin{equation}\label{res_img}
                    L_{r_{v}} = \left \| v_{i_1} - \hat{v}_{i_1} \right \|_1,
    \end{equation}
    where $v_{i_1}$ is the ground truth velocity field which is computed with corresponding small time-step simulation.
    
    \textit{Gradient Loss.}
    There is no guarantee that simply reducing the L1 loss of the velocity fields will also bring the derivative of the output velocity field closer to the ground truth \cite{kim2019deep}. To reduce the difference of the velocity gradients, we adopt the gradient loss $L_g$, which is defined as:
    \begin{equation}\label{res_img}
                    L_{g}= \left \| \nabla v_{i_1} - \nabla\hat{v}_{i_1} \right \|_1,
    \end{equation}
    where $\nabla$ denotes the gradient operation.
    
    \textit{Velocity Temporal Coherence Loss.}
    $\hat{v}_{i_1}$ should be temporally consistent with the initial $v_{i_0}$ to produce accurate results. Therefore, we adopt the temporal coherence loss \cite{oyj} to make the change from $v_{i_0}$ to $\hat{v}_{i_1}$ during time $t$ to be similar to the ground truth, the change from $v_{i_0}$ to $v_{i_1}$ during time $t$.
    The velocity temporal coherence loss $L_{t_v}$ is defined as:
    \begin{equation}
    L_{t_v}=\left \| \frac{d}{dt}
    \mathcal{C}(v_{i_0},{v}_{i_1}) - \frac{d}{dt}
    \mathcal{C}(v_{i_0},\hat{v}_{i_1}) \right \|_1,
    \end{equation}
    where $\mathcal{C}$ is a function that concatenates the two given inputs along the time axis.
    
    \textit{Velocity Perceptual Loss.}
    Perceptual loss is often used in image generation tasks to solve blurriness caused by pixel-wise L1 or L2 loss.
    It represents the visual similarity of two images by extracting the features of each image and calculating their difference.
    Using the pre-trained VGG \cite{vgg} network as a feature extractor, the perceptual loss makes the feature representation of the source image similar to that of the target image \cite{johnson2016perceptual}.
    We can apply this idea to velocity fields: by treating the x, y, and z components of a velocity field as the r, g, and b channels of an image, respectively, we can represent a velocity field as a heat map.
    We can then use the pre-trained VGG-16 network as a feature extractor to calculate the perceptual loss.
    It is possible to train a separate feature extractor using our data, but it is more efficient in terms of training overhead to use a pre-trained network as the feature extractor.
    
    To get clearer results by preserving the perceptual structure of the velocity fields, we add a perceptual loss for velocity fields $L_{p_{v}}$, which is defined as:
    \begin{equation}\label{res_img}
                    L_{p_{v}}= \left \| \phi(v_{i_1}) - \phi(\hat{v}_{i_1}) \right \|_2^2,
    \end{equation}
    where $\phi$ denotes the conv4\_3 feature of VGG-16 network \cite{vgg, jiang2018super} which is pre-trained with ImageNet \cite{deng2009imagenet}.
    
    Note that the $z$-values are zeros for 2D velocity fields.
    Applying perceptual loss to 3D velocity fields is described in detail in Section \ref{3ppl}.

    \subsubsection{Density Correction} \label{subsubsec:density_correction}
    As mentioned in Section \ref{sec:img_g_optical_flow}, it is difficult to model complex details when generating images directly with a DNN. Optical flow is often used to address this problem \cite{Liang_2017_ICCV, wang2018video, lu2021video}.
    Similarly, we adopt the concept of flow for correcting density fields, the final visible output of smoke simulations, to accurately visualize the complex details of smoke simulations.
    By appropriately advecting the imprecise density field computed by the large time-step simulation, we can obtain a result much closer to the density field of the corresponding small time-step simulation result, which we use as the ground truth.
    This advection may be represented by a flow from the large time-step simulation result to the ground truth.
    Thus our correction network first predicts the optical flow $F$ for density fields.
    Then we warp the input density field $\bar{\rho}_{i_1}$ using $F$.
    However, it is difficult to reconstruct the correct density field just by warping the large time-step simulation result. Simulating with larger time-steps accumulates larger errors during the calculation process. These errors may include density loss, which can be fatal to the accuracy of the warped result.
    Accordingly, our network thus also predicts a density field $\tilde{\rho}_{1}$ to supplement the warping result, and generates the weight $\alpha$ to determine the influence of the additional density (Figure \ref{fig:correction_network}).
    By fusing this additional density field with the warping result, our method can model both spatial and temporal information.
    The final output density $\hat{\rho}_{1}$ is calculated as:
    \begin{equation}\label{res_img}
        \hat{\rho}_{i_1}=\mathcal{W}(\bar{\rho}_{i_1}, F) + \alpha\tilde{\rho}_{i_1},
    \end{equation}
    where $\mathcal{W}$ denotes the warping function.
    The loss function to optimize the density correction $L_{\rho}$ is defined as:
        \begin{equation}\label{res_img}
                    L_{\rho}=\lambda_{r_{\rho}}L_{r_{\rho}} + \lambda_{I}L_{I} + \lambda_{p}L_{p_{\rho}} + \lambda_{t_\rho}L_{t_{\rho}}.
        \end{equation}
    $L_{r_{\rho}}$, $L_{I}$, $L_{p_{\rho}}$, and  $L_{t_{\rho}}$ denotes the reconstruction loss, interpolation loss, perceptual loss, and temporal coherence loss of the density fields, respectively.
    $\lambda_{r_{\rho}}$, $\lambda_{I}$, $ \lambda_{p}$, and $\lambda_{t_{\rho}}$ are the weights of each term.
    
    \setlength\itemsep{0.5em}
    \textit{Density Reconstruction Loss.}
    The reconstruction loss for density field $ L_{r_{\rho}}$ is defined as:
    \begin{equation}\label{res_img}
            L_{r_{\rho}}=\left \| \rho_{i_1} - \hat{\rho}_{i_1} \right \|_1
    \end{equation}
    where $\rho_{i_1}$ is the ground truth density field which is computed with the corresponding small time-step simulation.
    
    \textit{Interpolation Loss.}
    If the correction is done properly, the interpolation results between ${\rho}_{i_0}$ and $\hat{\rho}_{i_1}$ become closer to the ground truth intermediate smoke frames between ${\rho}_{i_0}$ and ${\rho}_{i_1}$.
    Using the interpolation network to be described in Section \ref{subsec:interpolation network}, we can obtain an intermediate density field between ${\rho}_{i_0}$ and ${\rho}_{i_1}$ at an arbitrary time $j \in (i_0, i_1)$.
    To make the entire sequence, including the correction and the interpolation results, closer to the ground truth, we propose the interpolation loss $L_{I}$ which is defined as:
    \begin{equation}\label{res_img}
            L_{I}=\left \| \rho_j - \hat{\rho}_j \right \|_1,
    \end{equation}
    where $\rho_j$ denotes the ground truth intermediate density field at frame index $j \in (i_0, i_1)$, and $\hat{\rho}_j$ denotes the intermediate density field obtained from the interpolation network at the $j$-th frame.
    
    \textit{Density Temporal Coherence Loss.}
    We again adopt the temporal coherence loss for density fields $L_{t_\rho}$ to model the changes from $\rho_{i_0}$ to $\hat{\rho}_{i_1}$ during time $t$ to be similar to the ground truth \cite{oyj}. This is defined as:
        \begin{equation}
        L_{t_\rho}=\left \| \frac{d}{dt}
        \mathcal{C}(\rho_{i_0},{\rho}_{i_1}) - \frac{d}{dt}
        \mathcal{C}(\rho_{i_0},\hat{\rho}_{i_1}) \right \|_1,
        \end{equation}
        where $\mathcal{C}$ is a function that concatenates the two given inputs along the time axis.

    \textit{Density Perceptual Loss.}
    As was applied to the velocity correction, the perceptual loss is also applied to the density correction.
    As the VGG-16 network we use to calculate perceptual loss requires three input channels, we triplicate the one-channel density field and channel-wise concatenate them before feeding it to the network. The perceptual loss for density fields $ L_{p_{\rho}}$ is defined as:
   \begin{equation}\label{res_img}
        L_{p_{\rho}}=\left \| \phi(\rho_{i_1}) - \phi(\hat{\rho}_{i_1}) \right \|_2^2.
    \end{equation}
    Applying perceptual loss to 3D density fields is described in detail in Section \ref{3ppl}. 
    
        \subsubsection{Applying Perceptual Loss to 3D Case}\label{3ppl}
        \begin{figure}[t]
        \centering
        \includegraphics[width=.45\linewidth]{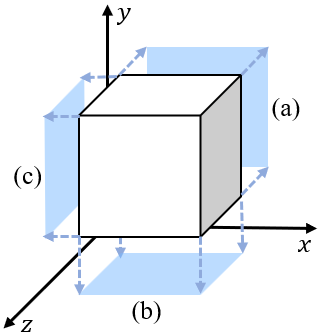}
        \caption{\label{fig:projection}
            Projection of a 3D tensor to a 2D plane. We randomly project the 3D velocity and density fields to one of \textnormal{(a)} $xy$,  \textnormal{(b)} $xz$, \textnormal{(c)} $yz$-planes, and then use the projection result as the input to the VGG-16 network.}
        \end{figure}
        We have computed the perceptual loss using the features from the pre-trained VGG-16 network in Section \ref{subsubsec:density_correction}.
        However, since the VGG-16 network receives 2D data as inputs, the same perceptual loss cannot be applied to 3D data.
        One solution would be designing and training a new 3D network and using the feature loss as a perceptual loss. This, however, is inefficient as it incurs the additional overhead of training a new 3D network.
        We therefore utilize the pre-trained VGG by transforming the 3D density and velocity fields into 2D.
        
        Figure \ref{fig:projection} shows the process of transforming a 3D tensor into 2D plane.
        We first get a planar view of the 3D density and velocity fields by randomly projecting them onto one of $xy$, $xz$, and $zy$-planes.
        The reason for random projection is to reduce training overhead. Projecting in all directions makes the loss too large, making the convergence slow. Projecting to a random direction for each iteration keeps the loss low, and repeated iteration eventually reduces the loss for all directions. The technique that focuses on a smaller part of the problem rather than the entire problem is widely used in synthesis tasks in deep learning \cite{wang2018video, chang2019free, vougioukas2020realistic}.
        
        We define the projection function as the mean of the values of the points along the orthogonal axis to the corresponding projection plane.
        For example, let $p(x, y, z)$ be the value at the point $(x, y, z)$ of the tensor. Then the value of each point on the projected 2D $xy$ plane, $P(x, y)$, is defined as:
        \begin{equation}
                P(x, y) = \frac{1}{N}\sum_{i=1}^N p(x, y, z_i),
        \end{equation}
        where $p(x, y, z_i)$ denotes the value at the $i$-th point of the tensor along the orthogonal axis $z$, and $N$ denotes the number of the points in the 3D tensor along the axis.
        
        Finally, we re-define the perceptual loss of 3D velocity fields as:
        \begin{equation}\label{res_img}
                    L_{p_{v}}=\left \| \phi(\psi(v_{i_1})) - \phi(\psi(\hat{v}_{i_1})) \right \|_2^2,
        \end{equation}
        where $\psi$ denotes the projection function. The density perceptual loss in 3D is also re-defined as:
        \begin{equation}\label{res_img}
                    L_{p_{\rho}}=\left \| \phi(\psi(\rho_{i_1})) - \phi(\psi(\hat{\rho}_{i_1})) \right \|_2^2.
        \end{equation}

\subsubsection{Implementation Details}
    \begin{figure}[t]
        \centering
        \includegraphics[width=1.\linewidth]{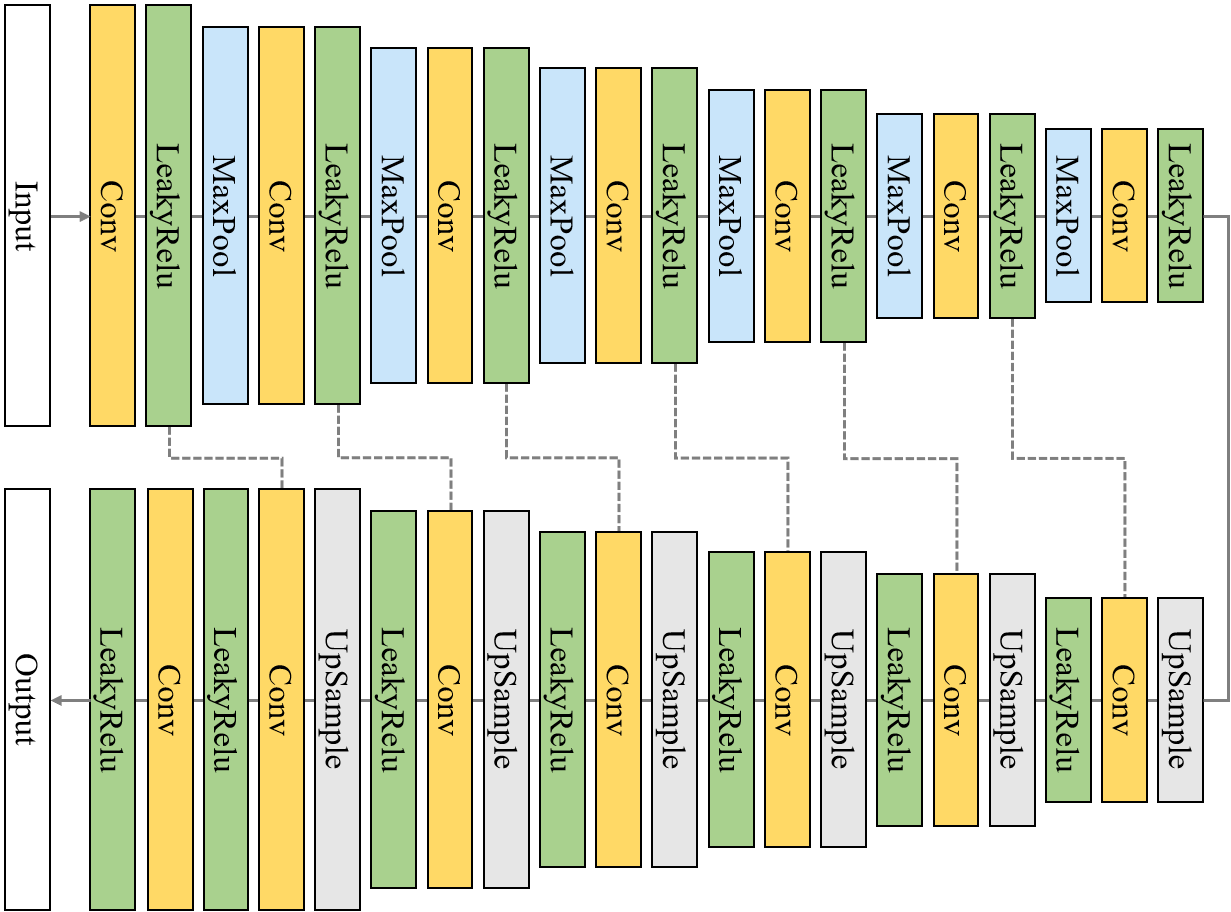}
        \caption{\label{fig:cn_architecture}
           Detailed structure of the correction network based on U-Net. The number of output channels increases to 16, 32, 64, 128, 256, and 512 within the encoding stage and decreases in reverse within the decoding stage.}
    \end{figure}
    
\begin{table*}[tb]
\caption{Statistics of training and test datasets}
\centering
  \begin{tabular}{p{1.8in} c c c c c c}
  \hline
  \multirow{2}{*}{Scenes} & Simulation & $\#$ of Simulations& $\#$ of Simulations& $\#$ of Frames & $\#$ of large-small pairs\\ 
  {} &  Grid Size & (train) & (test) & per Simulation  & in training set\\ \hline
    (a) Smoke2D Plume & 128$\times$128 & 1000 & 200 & 200 & 25000\\
    (b) Smoke2D $\&$ Fixed Circle & 256$\times$ 256 & 500 & 50 & 128 & 8000\\
    (c) Smoke3D Inflow & 80$\times$80$\times$80 & 100 & 20 & 200 & 2500\\ 
    (d) Smoke3D $\&$ Fixed Circle & 128$\times$128$\times$128 & 50 & 5 & 128 & 800\\\hline
  \end{tabular}
  \label{tab:data_statistics}
\end{table*}

    \setlength\itemsep{0.5em}
    \textit{Architecture Details.}
    The correction network is designed based on U-Net \cite{ronneberger2015u}. Figure \ref{fig:cn_architecture} shows the detailed structure of the network architecture.
    It consists of an encoder and a decoder, with skip-connections between the features of the same size in the encoder and decoder.
    The encoder and decoder each consist of seven blocks containing a convolutional layer and a Leaky-Relu layer.
    Each encoder block, except for the last block, has a max-pooling layer at the end.
    Each decoder block, except for the last block, has an up-sampling layer at the beginning.
    For the encoder blocks, the convolution kernel size decreases from seven in the first block, to five in the second block, then to three for the other blocks. For all the decoder blocks, the size of the convolutional kernel is three.
    The number of output channels increases to 16, 32, 64, 128, 256, 512, and 512 in the encoder blocks, and decreases in reverse in the decoder blocks.
   
   \textit{Hyperparameter Settings.}
    The network is implemented in python with PyTorch \cite{pytorch}.
    We use the Adam Optimizer \cite{kingma2014adam} with a learning rate of $10^{-3}$ and a batch size of 8 while training.
    When training 3D data, we use gradient accumulation to overcome the memory constraints.
    The weights of the loss terms are set as follows: 
    $\lambda_{r_v}$ and $\lambda_{t_v}$ are each set to 20.0, and  $\lambda_{r_{\rho}}$, $\lambda_{t_{\rho}}$, and $\lambda_{g}$ are each set to 50.0. $\lambda_{I}$ is set to 1.0.
    $\lambda_{p}$ is set to $10^{-1}$ for 2D data and $10^{-6}$ for 3D data.
    
\subsection{Interpolation Network}\label{subsec:interpolation network}
    The interpolation network interpolates between $\rho_{i_0}$ and $\hat{{\rho}}_1$ to increase the frame rate.
    We adopt the two-step temporal interpolation model for smoke frames using optical flow, since it is the fastest model to generate the intermediate frames \cite{oyj}.
    
    The model in the first step interpolates between ${\rho}_{i_0}$ and $\hat{{\rho}}_{i_1}$.
    Then the second-step model interpolates between $\hat{{\rho}}_{i_1}$ and the forward advection result of ${\rho}_{i_0}$ using $v_{i_0}$.
    When training the interpolation network, we add a perceptual loss term to the original loss function proposed in the previous work \cite{oyj}, and also apply the projection technique introduced in Section \ref{3ppl} for 3D density fields.
    
    The forward advection result from the second step is used to produce clearer and more accurate interpolation results.
    At the training phase of the correction network, where we know the exact velocity vector $v_{i_0}$, we use the second-step interpolation result for computing interpolation loss $L_I$.
    However, in the testing phase, the error of the simulated velocity field increases as the simulation rollout continues.
    If the velocity field deviates too far from the ground truth, the forward advection result becomes inaccurate. In that case, the result from interpolating between $\hat{{\rho}}_{i_1}$ and the defective forward advection may become less accurate than the result from the first step.
    Therefore, in the first half of the simulation, where the velocity field is relatively closer to the ground truth, we use the second-step interpolation result. In the latter half, we use the first step interpolation result.

\section{Experiments}
\label{sec:experiments}
This section explains the details of the training and test phase, then evaluates the proposed method.
In the training phase, the correction network is trained to correct only the results after a single-step simulation using a large time-step.
In the test phase, we evaluate our method in two ways.
First, we evaluate the corrected result of the single-step simulation with a large time-step.
Second, we conduct a long-term simulation using our method, then compare the results with the corresponding small time-step simulation results.
If the model is trained to generalize complex fluid dynamics, it should be able to generate desirable results for long-term simulations, even if it is trained only with single-step simulations.

The long-term simulation using our method is described in Algorithm 1.
In the beginning, a physics-based solver calculates a single step of a large time-step simulation from the initial state.
This result is corrected by the correction network, and then the interpolation network interpolates between the correction result and the initial state.
The corrected output is fed to the solver as the input for the next iteration, and the process is continued until the frame index reaches the final index $N$.

\begin{algorithm}[t]
\label{algo}
\renewcommand{\algorithmicrequire}{\textbf{Input:}}
\renewcommand{\algorithmicensure}{\textbf{Output:}}
\begin{algorithmic}[1]
\Require{initial state at frame index 0
        \par $\Delta t_s$, small time-step
        \par $\Delta t_L$, large time-step
        \par $N$, final frame index}
\Ensure{$\rho_{i_1}, ..., \rho_N$, density fields from index 1 to N}
\State{$k \gets \Delta t_L / \Delta t_s$}
\State{$i_0 \gets 0$}
\While{$i_0 < N$}
    \State{$i_1 \gets i_0 + k$}
    \State{$\rho_{i_1}$, $v_{i_1} \gets Solver({\rho}_{i_0}, {v}_{i_0}, \Delta t_L)$}
    \State{$\rho_{i_1}$, $v_{i_1} \gets Correct({\rho}_{i_0}, {v}_{i_0}, \rho_{i_1}$, $v_{i_1})$}
    \For{$j$ = 1, 2, ..., $k-1$}
        \State {$\rho_{i_0+j} \gets Interpolate(\rho_{i_0}, \rho_{i_1}, v_{i_0}, j)$}
     \EndFor
    \State{$i_0 \gets i_0 + k$}
\EndWhile
\end{algorithmic}
\caption{Simulation using our method}
\end{algorithm}

 \begin{figure}[t]
    \centering
    \includegraphics[width=.94\linewidth]{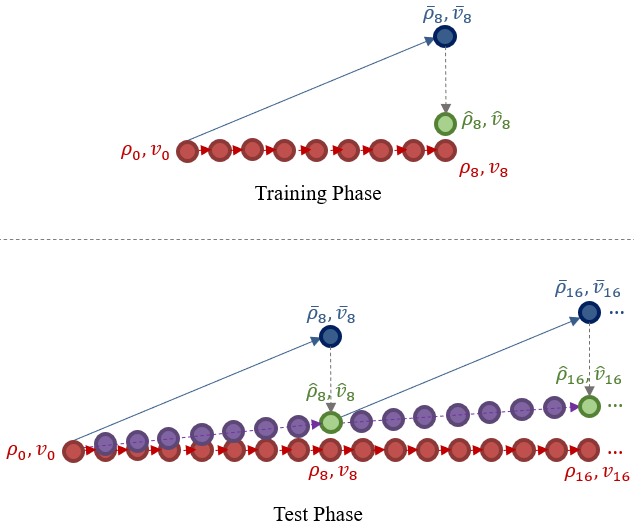}
    \caption{\label{fig:train_test}
       Illustration of the training phase and test phase.
       In the training phase, the model is trained to generate the corrected output (green) of the single large time-step simulation results (blue) to be close to the corresponding target small time-step simulation results (red). The simulation is not continued after the correction.
       In the test phase, after a single large step correction, the process continues to generate more than 100 frames. We check whether the correction results and intermediate frames (purple) are similar to the target.
       Note that the initial state $\rho_0, v_0$ of the training and test phase are always different.
       } 
    \end{figure}

The datasets are generated using Mantaflow \cite{mantaflow}, an open-source physics-based CPU solver for fluid simulation. The small time-step is set to 0.5, and the large time-step is set to 4.0.
We first generate initial states $\rho_0, v_0$ with random seeds, then rollout 128 or 200 steps in the small time-step.
These simulation results are separated into training and test data.
To make small-large time-step simulation result pairs for training data, we conduct one large time-step simulation from every $8n$\textsuperscript{th} frame ($n=0, 1, 2, ...$) of small time-step simulations.
For test data, we use the entire long-term simulation results, which are not included in the training data.
Table \ref{tab:data_statistics} shows the detailed statistics of the datasets.

Fig \ref{fig:train_test} illustrates the difference between the training and test phases using Algorithm 1 in detail.
$\rho_i, v_i$ denotes the density and velocity field of the $i$-th frame, respectively.
In the training phase, the network is trained to correct only the results after a single-step simulation. The simulation is not continued after the correction.
On the other hand, in the test phase, the simulation is continued at the correction value after the first large step.
Note that the initial state $\rho_0, v_0$ in the training phase and the test phase are always different values since our test dataset consists only of the data not included in the training set.

The proposed framework is trained on a single NVIDIA RTX 3090 GPU, with an i7-10700 CPU at 2.9GHz, and tested on a single NVIDIA RTX 3090 GPU with a 5600X CPU at 3.7GHz.
We trained the proposed model and its comparatives for 200 epochs, and saved the best model with the lowest validation loss among all training epochs.

\subsection{Comparison of corrected results and ground truth}
    In this section, we evaluate the effectiveness of the correction network using the large-small pairs of the test data.
    Table \ref{tab:correction_rate} shows the average mean squared error (MSE) of the density fields and velocity fields of large time-step simulation results on each dataset.
    Table \ref{tab:correction_rate}(a), (b), (c), and (d) correspond to each dataset of Table \ref{tab:data_statistics}.
    The first and second columns of each field denote the average MSE before and after correction. The third columns show the error reduction rates.
    For 2D datasets, the errors of the density fields are reduced by more than 80\%, and the errors of the velocity fields are reduced by more than 90\% (Table \ref{tab:correction_rate}(a), (b)).
    For 3D dataset with low-resolution, the density and velocity errors are reduced by about 80\% and 79\%, respectively (Table \ref{tab:correction_rate}(c)).
    For 3D dataset with high-resolution, each error is reduced by about 42\% and 69\%, respectively  (Table \ref{tab:correction_rate}(d)).
    The possible reason that the error reduction rate of the density fields is comparatively low in Table \ref{tab:correction_rate}(d) is the small initial density difference between the large and small time-step results.
    Since the errors were already very small for density fields even without correction (0.039 $\times 10^{-2}$), the model would have put more effort into correcting the errors of the velocity field (2.575 $\times 10^{-2}$), which were initially 66 times greater than those of the density field.
    
\begin{figure}[t]
    \centering
    \includegraphics[width=.9\linewidth]{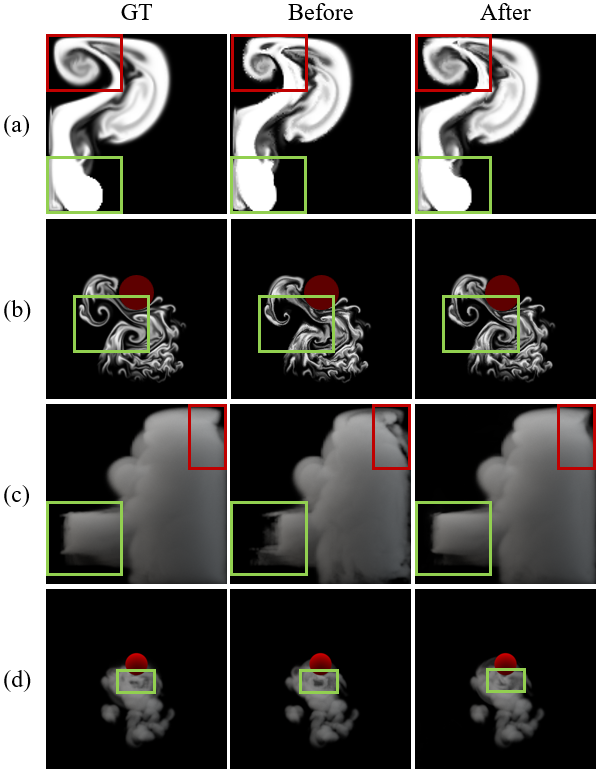}
    \caption{\label{fig:correction_samples}
       Single-step correction results of samples from each dataset.
       The first column shows the ground truth, and the second and third columns show the results of the large time-step simulation before and after correction, respectively.
       Rows \textnormal{(a)}, \textnormal{(b)}, \textnormal{(c)}, and \textnormal{(d)} indicates the datasets in Table \ref{tab:data_statistics}, where red circles in (b) and (d) are obstacles.} 
    \end{figure}

    Figure \ref{fig:correction_samples} shows the single-step density correction results of the samples from each dataset, where (a), (b), (c) and (d) indicate the datasets in Table \ref{tab:data_statistics}.
    The first, second, and third columns of Figure \ref{fig:correction_samples} show the density field of the ground truth, the large time-step simulation result before correction, and the corrected result, respectively.
    The red circles in Figure \ref{fig:correction_samples}(b) and (d) are fixed obstacles in the simulation scenes.
    Major differences in the simulations are highlighted by the colored boxes, showing the effectiveness of the correction network.
    In particular, the red box of Figure \ref{fig:correction_samples}(a) and the green box of Figure \ref{fig:correction_samples}(b) show that the correction network makes vortex-like structures become more similar to the ground truth.
    For the boxes in Figure \ref{fig:correction_samples}(c) and (d), while results in the second column are empty in some parts, the corrected results properly fill in the empty spaces.
    To summarize, the qualitative and quantitative results show that the proposed correction network effectively corrects the large time-step simulation results.
    
    \begin{table}[t]
    \caption{MSE of density and velocity fields of large time-step simulation results before and after correction}
    \centering
    \resizebox{\columnwidth}{!}{
    \begin{tabular}{c|ccc|ccc}
    \noalign{\smallskip}\noalign{\smallskip}\hline\hline
    \multirow{3}{*}{} & \multicolumn{3}{c|}{Density} & \multicolumn{3}{c}{Velocity} \\
    \cline{2-7}
          & Before & After & Reduced & Before & After & Reduced \\
          &  ($\times$100) & ($\times$100) & (\%) &  ($\times$100) & ($\times$100) & (\%) \\
    \hline
     (a) & 1.459 & 0.100 & 93.09 & 2.292 & 0.124 & 94.54\\
     (b) & 0.540 & 0.105 & 80.53 & 4.137 & 0.370 & 91.03\\
     (c) & 1.213 & 0.236 & 80.51 & 4.951 & 1.013 & 79.53\\
     (d) & 0.039 & 0.020 & 42.27 & 2.575 & 0.781 & 69.64\\
    \hline
    \hline
    \end{tabular}
    }
    \label{tab:correction_rate}
    \end{table}
    
\subsection{Comparison with previous DNN-based methods}
    In this section, we compare the quality of our results with previous DNN-based methods that directly generate the simulation results: DeepFluids (DF) \cite{kim2019deep}, an LSTM-based prediction model using Latent Space Subdivision (LSS) \cite{wiewel2020latent}, and the naive Two-Step Interpolation model (TSI) \cite{oyj}.
    The comparison target is the density field, which is the final visible output of the simulation.
    The comparative models are trained with the same GPU computing power as used in the training of our model.
    For DF, we parameterized the data by time; we first decode the smoke source and the time-step information to obtain the velocity field, then forward-advect the density field using the velocity field.
    For LSS, we directly use the decoded density field.
    According to Wiewel et al. \cite{wiewel2020latent}, it is possible to get these density fields without encoding at every step.
    However, we experimentally found that encoding and decoding at every step produce better results, so we chose this method.
    For TSI, we use the interpolation results between the raw large time-step simulation results without correction.
    
    \begin{figure}[t]
        \centering
        \includegraphics[width=.9\linewidth]{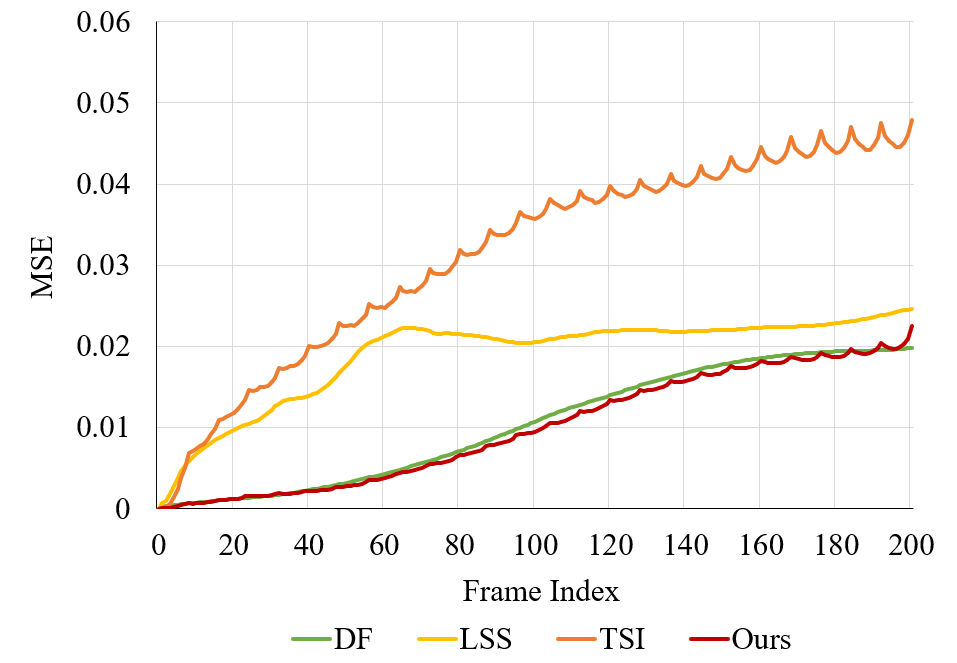}
        \caption{\label{fig:mse_plot}
           Average MSE of each method when generating density fields at each frame. Our approach shows the lowest MSE for all frame indexes.}
    \end{figure}
    
    \begin{figure*}[t]
      \centering
      \mbox{} \hfill
      \includegraphics[width=.7\linewidth]{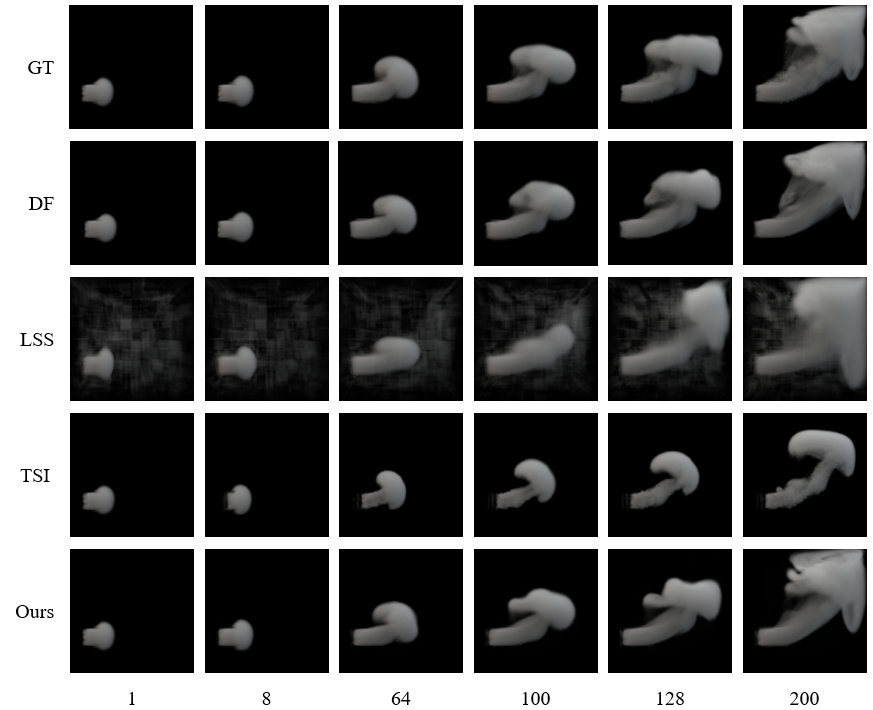}
      \hfill \mbox{}
      \caption{\label{fig:comparison_80^3}
      Density field samples of 1\textsuperscript{st}, 8\textsuperscript{th}, 64\textsuperscript{th}, 100\textsuperscript{th}, 128\textsuperscript{th}, and 200\textsuperscript{th} step from the \textit{Smoke3D Plume} dataset. The numbers below indicate the frame index. The top row is the ground truth. The second, third, and fourth rows are the results produced by DF\cite{kim2019deep}, LSS\cite{wiewel2020latent}, and TSI\cite{oyj}, respectively. The last row shows our results, which are the closest to the ground truth.}
    \end{figure*}
    
    Figure \ref{fig:mse_plot} plots the average MSE of density fields at each of the 200 frames of \textit{Smoke3D Inflow} dataset.
    In all methods, the error is small at the beginning, but increases as the simulation continues.
    The graph of TSI exhibits a wave-like pattern with an 8-step cycle, where the peaks correspond to the interpolation inputs. These correspond to the results of the physics-based solver simulated with a large time-step.
    Since it is trained to generate the interpolation result close to ground truth, the interpolation results show lower MSE compared than the inputs.
    The graph of our model also has a similar pattern to TSI, but the overall MSE is significantly lower.
    Even compared to other models, our model has the smallest error for the entire range.
    The average error of the 200 points is 0.0104, 0.0190, 0.0315, and 0.0099, for DF, LSS, TSI, and ours, respectively.
    The proposed method is on average 2.04 times accurate than other methods.
    
    Figure \ref{fig:comparison_80^3} shows the qualitative comparison of the generated samples.
    The horizontal axis represents the frame index, while the vertical axis represents each method, where the top row is the ground truth.
    Using forward advection and parameterization by time, DF decently reconstructs the test data without artifacts.
    On the other hand, the results of LSS show that directly generating density fields without additional advection is prone to producing artifacts.
    In the case of TSI, since their interpolation input frames (8\textsuperscript{th}, 64\textsuperscript{th}, 128\textsuperscript{th}, and 200\textsuperscript{th} frames) are different from the ground truth, their interpolation results (1\textsuperscript{st} and 100\textsuperscript{th} frames) also deviate from the ground truth.
    On the other hand, our method effectively corrects the interpolation input frames and produces results closer to the ground truth. Unlike LSS, our method does not produce artifacts even though it generates density fields without advection.
    Our method also produces results similar to the ground truth even up to the 200\textsuperscript{th} frame despite the training data consisting only of single-step simulations.
    
    Through the quantitative and qualitative evaluations, we have found that our method produces density fields without extra advection.
    We also have found that our method produces desirable long-term rollout results even though the training data consisted only of single-step simulations.
    We can infer that our model does not simply memorize the density or velocity field of single-step simulation but is generalized to model the fluid dynamics well.

\subsection{Performance Analysis}
    \begin{table}[t]
    \caption{Average computation time of each method \\
    when generating 200 frames of smoke scenes} 
    \label{tab:simultime}
    \centering
      \resizebox{\columnwidth}{!}{
      \begin{tabularx}{\columnwidth}{p{0.34\linewidth} *{3}{>{\centering\arraybackslash}X}}
      \hline
      \multirow{2}{*}{Method}   & Simulation       & Inference  & Total            \\
                                & time (s)         & time (s)   & time (s)         \\ \hline
        Mantaflow \cite{mantaflow}           & 59.66            & -          & 59.66            \\
        DF \cite{kim2019deep}              & 2.34               & 20.73         & 23.07            \\
        LSS \cite{wiewel2020latent}                     & -               & {21.00}         & {21.00}               \\
        Ours (Correction Only)       & -               & {00.67}         & {00.67} \\
        Ours (Full)                    &8.49               &9.12         & \textbf{17.61}\\ 
      \hline
      \end{tabularx}
      }
    \end{table}
    We also compare the frame generation speed of our method with the previous DNN-based methods and the physics-based solver, Mantaflow.
    We used batch size 1 for all DNN-based methods, including ours.
    Table \ref{tab:simultime} shows the average time it takes to generate 200 frames from \textit{Smoke3D Inflow} dataset.
    Simulation time indicates the time used by Mantaflow, which is run on CPU, and inference time indicates the time used by the models, which are run on GPU.
    For DF, simulation time refers to the density advection time.
    LSS requires the model inference time only, since it directly uses the decoded density fields.
    The total computation time of the proposed method consists of the large time-step simulation time and model inference time, where the model inference time includes correction time and interpolation time.
    Although our method has the longest simulation time among the three DNN-based methods, it is ahead in the total time because the inference time of our method is less than half that of other methods.
    The proposed correction method takes 0.67 seconds on average to correct 25 frames.
    While effectively correcting the large time-step simulation output, the computation time of the correction does not significantly affect the overall inference time, making the inference time bound to the interpolation time.
    While it is already slightly faster than previous methods, our method has the potential to embrace parallel computation during interpolation (line 7$\sim$9 in Algorithm 1) to make it even faster \cite{oyj}.

\subsection{Ablation studies}
    \subsubsection{Effectiveness of combining optical flow based warping and density generation}
    We first investigate the effectiveness of combining the warping result with the predicted flow($\bar{\rho}_{i_1}$) and the predicted density field($\tilde{\rho}_{i_1}$).
    We train two different versions of the correction network on 2D datasets: one in which the density is removed, and the other in which the flow is removed from the prediction target.
    As we use 2D datasets, we measure SSIM and LPIPS \cite{chu2020tecoGAN}, which are applicable only to 2D data, in addition to MSE.
    SSIM is an index indicating structural similarity between images, and LPIPS is a perceptual similarity measure based on DNNs.
    Higher SSIM and lower LPIPS indicate that the output is more similar to the ground truth.
    
\begin{figure}[t]
      \centering
      \includegraphics[width=1.\linewidth]{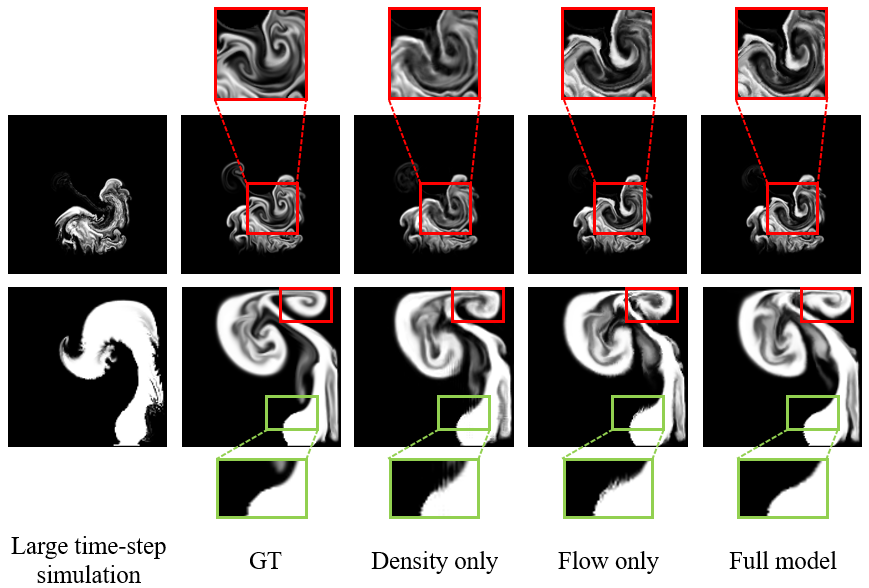}
      \caption{\label{fig:ablation_flowdenvel}
      Samples of output density fields generated by each model with removed prediction target and original full model. GT denotes the ground truth.}
    \end{figure}

    Table \ref{tab:ablation_flowdenvel} shows the average MSE, SSIM, and LPIPS of generating 200 and 128 frames of the test set of \textit{Smoke2D Plume} and \textit{Smoke2D \& Fixed Circle} dataset, respectively.
    On \textit{Smoke2D Plume} dataset, the original full model showed the lowest MSE and LPIPS, and the highest SSIM (Table \ref{tab:ablation_flowdenvel}(a)).
    For \textit{Smoke2d \& Fixed Circle} dataset, SSIM and LPIPS were the lowest in the original full model, but the MSE was lowest in the density-only model (Table \ref{tab:ablation_flowdenvel}(b)).
    However, low MSE does not always guarantee high quality.
    Figure \ref{fig:ablation_flowdenvel} shows the samples from each dataset for qualitative assessment.
    The leftmost column is the simulation results calculated with large time-step, and the second column is the corresponding ground truth calculated with small time-step.
    The other three columns are the density fields generated by each model.
    The model which is trained to predict only the density fields tends to be blurry due to oversmoothing.
    In addition, artifacts easily occur when we do not use the full model.
    The full model best preserves the vortex shape in the red boxes and the bulge in the green boxes, among the models.
    The results indicate that combining $\bar{\rho}_{1}$ and $\tilde{\rho}_{1}$ leads to the most desirable results.
    
    \begin{table}[t]
    \caption{Evaluation on generated frames of 
    \\(a) \textit{Smoke2D Plume} dataset and (b) \textit{Smoke2D \& Fixed Circle} dataset}
    \centering
    \begin{tabularx}{\linewidth}{p{0.05\columnwidth}|p{0.2\columnwidth}|*{3}{>{\centering\arraybackslash}X}}
    \noalign{\smallskip}\noalign{\smallskip}\hline\hline
    \multicolumn{2}{p{0.1\linewidth}|}{} & MSE($\downarrow$) & \multirow{2}{*}{SSIM($\uparrow$)} & LPIPS($\downarrow$) \\
    \multicolumn{2}{p{0.1\linewidth}|}{} & ($\times$1000) & {} & ($\times$1000) \\
    \hline
    \multirow{3}{*}{(a)} & Density Only  & {19.27}  &{0.8163} & {8.202} \\
    & Flow Only & {17.44}  &{0.8250} & {8.084} \\
    & Ours  & \textbf{17.17} &\textbf{0.8291} & \textbf{7.652} \\
    \hline
    \multirow{3}{*}{(b)} & Density Only  & \textbf{7.296}  &{0.8175} & {10.866} \\
    & Flow Only & {7.328}  &{0.8250} & {9.779} \\
    & Ours  & {7.393} &\textbf{0.8254} & \textbf{9.769} \\
    \hline
    \hline
    \end{tabularx}
    \label{tab:ablation_flowdenvel}
    \end{table}

    \begin{figure}[t]
          \centering
          \includegraphics[width=.8\linewidth]{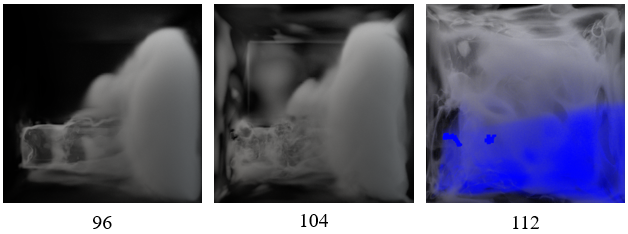}
          \caption{\label{fig:wopercep}
          Frames from a sample output with density divergence. The numbers below the images indicate the frame index. The density field diverges in some samples when the model is trained without perceptual loss and temporal coherence loss. The blue parts show where values diverge to infinity, causing rendering to fail.}
        \end{figure}
    
    \subsubsection{Effectiveness of loss terms}
\begin{figure}[t]
          \centering
          \includegraphics[width=.9\linewidth]{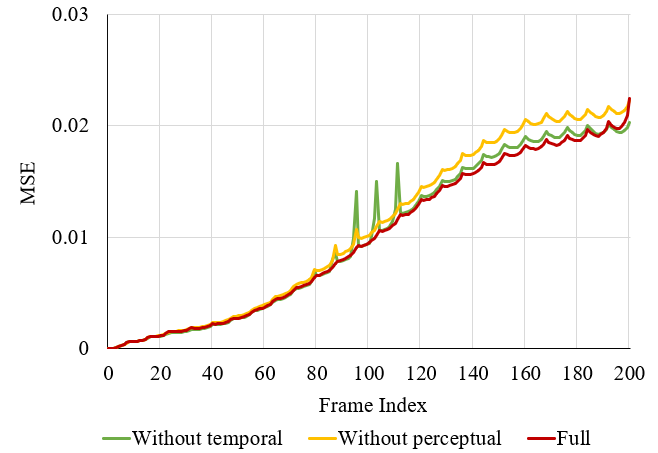}
          \caption{\label{fig:ablatation_temporal_percep}
          Average MSE of density fields at each step for each case of the ablation study. 'Without temporal' and 'Without perceptual' indicate the model trained without temporal coherence loss and perceptual loss, respectively. 'Full' indicates the full model trained with both of the loss terms.}
        \end{figure}

        We also investigate the effectiveness of using perceptual loss with the projected 3D tensors and temporal coherence loss.
        We train four different versions of the correction network on \textit{Smoke3D Inflow} dataset, removing or adding each of the two loss terms. The version with both loss terms is the original network.
        When both the perceptual loss and temporal coherence loss are removed, the density often diverges before completing 200 steps, making it impossible for the physics-based solver to continue the simulation (Figure \ref{fig:wopercep}).
        Therefore, the average MSE for each step could not be calculated due to the presence of infinite values.
        We thus exclude this version of the model from the subsequent experiments.
        Figure \ref{fig:ablatation_temporal_percep} plots the average MSE of density fields at each of the 200 frames, using each version of the model.
        The errors are similar at first, but differences become visible after the 80\textsuperscript{th} frame.
        The error is largest when the perceptual loss is removed, which infers that the perceptual loss has a greater effect than the temporal coherence loss.
        This indicates that applying perceptual loss with the projected 3D tensors drives the result closer to the ground truth.
        When the temporal coherence loss is removed, three notable peaks appear. These peaks, corresponding to the 95\textsuperscript{th}, 103\textsuperscript{rd}, and 111\textsuperscript{th} frames, are where artifacts occurred during the interpolation process.
        Since the interpolation network is sensitive to temporal coherence, artifacts may occur when the two inputs are not temporally coherent.
        The result indicates that the temporal coherence loss helps the model to generate temporally coherent outputs.
        The average of the 200 points is  0.0102, 0.0108, and 0.0099 for the model without temporal coherence loss, without perceptual loss, and the full model, respectively.
        Although it experiences a rapid increase after the 190\textsuperscript{th} frame, the MSE of the full model is still the lowest on average.
        We can conclude that using both of the losses helps to reduce the error accumulation for most cases.
        
\section{Discussions \& Future Works}
\label{sec:discussion}
As our results show, the correction we apply is effective, but the errors are not entirely removed.
Since the outputs of the correction network are used as inputs for subsequent calculations, errors gradually accumulate in our simulation.
One possible way to mitigate error accumulation is to add a network that corrects the long-term simulation results in addition to our existing correction network.
Another solution could be to train the model to correct not only the single-step simulations from the ground truth, but also the simulation results calculated from the frames with errors.
This can be done by including the outputs of the correction network to training data, and training the model to correct them to be closer to the ground truth.
Applying these solutions could reduce the amount of errors accumulated, but make the model heavier or increase the training overhead.
Weighing these costs and benefits would be an intriguing future study.

\begin{figure}[tbp]
          \centering
          \includegraphics[width=.88\linewidth]{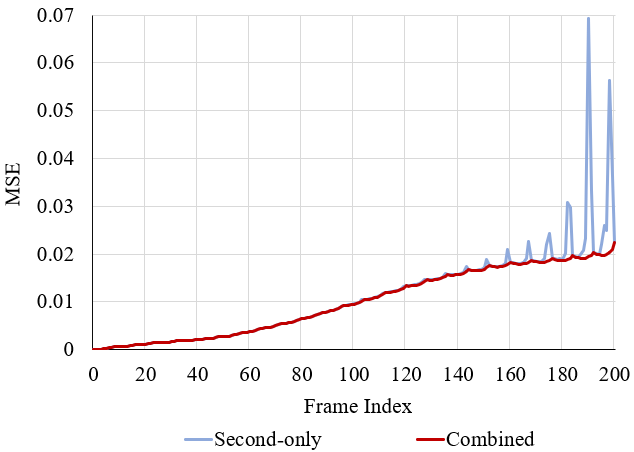}
          \caption{\label{fig:discussion_interpolation}
          Average MSE using different interpolation strategies. 'Second-only' indicates when only the second-step-model is used in the interpolation. 'Combined' indicates when the first-step-model is used in addition.}
\end{figure}

Modifications in the perceptual loss may also help improve the accuracy of our model. Our model currently uses the pre-trained VGG network as the feature extractor, which outputs general-purpose features. Instead, it could incorporate a new autoencoder trained with the velocity or density field data as the feature extractor. The features extracted by this autoencoder would be more task-specific than those extracted by the VGG network. Examining how much this task-specific feature affects the output quality would be interesting future work.

Our method also suffers from the inherent limitation of the interpolation method used.
The model we used for interpolation, TSI, follows a two-step process as explained in Section \ref{subsec:interpolation network}.
The second-step model of TSI uses the velocity field in the interpolation process.
However, as the simulation continues, errors accumulate in the velocity field. When this error exceeds a certain threshold, the second-step model outputs defective intermediate frames.
Figure \ref{fig:discussion_interpolation} shows the comparison of MSE when interpolation is performed only with the second-step model and when interpolation is performed using the first-step-model after the 100\textsuperscript{th} step.
The striking peaks in the graph indicate where artifacts occur.
If our correction network is combined with a better interpolation method, better results can be expected.

\begin{table}[t]
\caption{MSE of density and velocity fields of large time-step simulation results before and after correction, with varying time-step settings of \textit{Smoke2D Plume} dataset.}
    \centering
    \resizebox{\columnwidth}{!}{
    \begin{tabular}{c|cc|cc|cc}
    \noalign{\smallskip}\noalign{\smallskip}\hline\hline
    \multirow{3}{*}{} & \multicolumn{2}{c|}{} & \multicolumn{2}{c|}{Density} & \multicolumn{2}{c}{Velocity} \\
    \cline{2-7}
          {} & Small & Large & Before & After & Before & After \\
          {} & time-step & time-step &  ($\times$100) & ($\times$100) &  ($\times$100) & ($\times$100) \\
    \hline
     (a) & 0.5 & 8.0 & {43.469} & {5.278} & {100.410} & {5.412} \\
     (b) & 0.25 & 4.0 & {8.288} & {0.213} & {3.588} & {0.174} \\
     (c) & 0.5 &  4.0 & {1.459} & {0.100} & {2.292} & {0.124} \\
    \hline
    \hline
    \end{tabular}
    }
    \label{tab:discussion}
\end{table}

An important factor to take into account when using our method is the appropriate size of the time-steps.
Table \ref{tab:discussion} shows the error reduction rate of the correction network trained on \textit{Smoke2D Plume} dataset, with different time-step configurations.
First, consider the ratio of the large time-step size to the small time-step size.
While this ratio is the same for Table \ref{tab:discussion}(a) and (b), the absolute size of the large time-step is larger in (a).
In fact, the MSE after correction of Table \ref{tab:discussion}(a) is more than 24 times greater than that of Table \ref{tab:discussion}(b), although the large time-step size of (a) is only twice that of (b).
We can conclude that the absolute size of the large time-step contributes more to the model's accuracy than the ratio between the large and small time-steps.
Therefore, setting the large time-step to smaller values leads to more accurate results.
Now consider Table \ref{tab:discussion}(a) and (c). 
The MSE after the correction is lower in (c), supporting the claim that smaller large time-step sizes induce smaller errors.
However, naively choosing low values of large time-steps is undesirable.
For example, suppose that the large time-step size in Table \ref{tab:discussion}(c) is lowered to 2.0. This would reduce the MSE but would severely impact the generation time, because the physics-based solver would need to be used twice as frequently.
We should carefully consider this trade-off between computation time and accuracy.
While the configuration in Table \ref{tab:discussion}(c) is the best of the three configurations in the table, finding the optimal pair of time-step settings may also be interesting future work.

Another method worth considering in the accuracy-time tradeoff is the conservative advection \cite{10.5555/2422356.2422373, 10.1145/2019406.2019419}, which improves the traditional non-conservative semi-Lagrangian method.
However, conservative advection requires some additional operations to the previous advection method, which can be time-consuming. If time constraint is not an issue, conservative advection can be used in place of the pure semi-Lagrangian advection with a large time-step in our methodology to obtain better results.
\section{Conclusion}
\label{sec:conclusion}
We have proposed CIMS: a novel correction-interpolation method for efficient smoke simulation using DNN.
Our method can generate high frame rate smoke frames at low costs by correcting large time-step simulation results to be closer to the small time-step simulation results and interpolating between them.
Even though our correction method does not perform any additional advection, it effectively corrects the large time-step simulation without forming artifacts.
For velocity fields, the error is reduced by more than 90\% in 2D datasets, and 80\% in 3D datasets after correction.
Among the four datasets, the error of the density field is reduced by more than 80\% in the three datasets, and more than 40\% in the other dataset.
We have shown that the proposed method is more accurate than previous DNN-based methods, especially overcoming the limitation of the previous temporal interpolation method.
We have also found that the time required for correction is only about 3.8\% of the total time.
The results of our ablations studies indicate that the combination of flow-based warping and density prediction produces detailed results without artifacts. The use of both perceptual loss of projected 3D tensors and temporal coherence loss is also effective for reducing errors.


%

\ifCLASSOPTIONcompsoc
  \section*{Acknowledgments}
\else
  \section*{Acknowledgment}
\fi

This research was supported by the National Research Foundation of Korea (NRF) grant funded by the Korea government (MSIT). (No. NRF-2020R1A2C2014622)

\ifCLASSOPTIONcaptionsoff
  \newpage
\fi



\bibliographystyle{IEEEtran}
\bibliography{mbody.bbl}

\begin{thebibliography}{10}
\providecommand{\url}[1]{#1}
\csname url@samestyle\endcsname
\providecommand{\newblock}{\relax}
\providecommand{\bibinfo}[2]{#2}
\providecommand{\BIBentrySTDinterwordspacing}{\spaceskip=0pt\relax}
\providecommand{\BIBentryALTinterwordstretchfactor}{4}
\providecommand{\BIBentryALTinterwordspacing}{\spaceskip=\fontdimen2\font plus
\BIBentryALTinterwordstretchfactor\fontdimen3\font minus
  \fontdimen4\font\relax}
\providecommand{\BIBforeignlanguage}[2]{{%
\expandafter\ifx\csname l@#1\endcsname\relax
\typeout{** WARNING: IEEEtran.bst: No hyphenation pattern has been}%
\typeout{** loaded for the language `#1'. Using the pattern for}%
\typeout{** the default language instead.}%
\else
\language=\csname l@#1\endcsname
\fi
#2}}
\providecommand{\BIBdecl}{\relax}
\BIBdecl

\bibitem{4015404}
B.~Kim, Y.~Liu, I.~Llamas, and J.~Rossignac, ``Advections with significantly
  reduced dissipation and diffusion,'' \emph{IEEE Transactions on Visualization
  and Computer Graphics}, vol.~13, no.~1, pp. 135--144, 2007.

\bibitem{solenthaler2011two}
B.~Solenthaler and M.~Gross, ``Two-scale particle simulation,'' \emph{ACM
  Transactions on Graphics}, vol.~30, no.~4, pp. 1--8, 2011.

\bibitem{golas2012large}
A.~Golas, R.~Narain, J.~Sewall, P.~Krajcevski, P.~Dubey, and M.~Lin,
  ``Large-scale fluid simulation using velocity-vorticity domain
  decomposition,'' \emph{ACM Transactions on Graphics}, vol.~31, no.~6, pp.
  1--9, 2012.

\bibitem{7127055}
Y.~Yang, X.~Yang, and S.~Yang, ``A fast iterated orthogonal projection
  framework for smoke simulation,'' \emph{IEEE Transactions on Visualization
  and Computer Graphics}, vol.~22, no.~5, pp. 1492--1502, 2016.

\bibitem{ando2015dimension}
R.~Ando, N.~Thürey, and C.~Wojtan, ``A dimension-reduced pressure solver for
  liquid simulations,'' \emph{Computer Graphics Forum}, vol.~34, no.~2, pp.
  473--480, 2015.

\bibitem{yan2016multiphase}
X.~Yan, Y.-T. Jiang, C.-F. Li, R.~R. Martin, and S.-M. Hu, ``Multiphase sph
  simulation for interactive fluids and solids,'' \emph{ACM Transactions on
  Graphics}, vol.~35, no.~4, pp. 1--11, 2016.

\bibitem{7845705}
X.~Liao, W.~Si, Z.~Yuan, H.~Sun, J.~Qin, Q.~Wang, and P.-A. Heng, ``Animating
  wall-bounded turbulent smoke via filament-mesh particle-particle method,''
  \emph{IEEE Transactions on Visualization and Computer Graphics}, vol.~24,
  no.~3, pp. 1260--1273, 2018.

\bibitem{sato2021stream}
S.~SATO, Y.~DOBASHI, and T.~KIM, ``Stream-guided smoke simulations,'' 2021.

\bibitem{tompson2017accelerating}
J.~Tompson, K.~Schlachter, P.~Sprechmann, and K.~Perlin, ``Accelerating
  eulerian fluid simulation with convolutional networks,'' in
  \emph{International Conference on Machine Learning}.\hskip 1em plus 0.5em
  minus 0.4em\relax PMLR, 2017, pp. 3424--3433.

\bibitem{xie2018tempogan}
Y.~Xie, E.~Franz, M.~Chu, and N.~Thuerey, ``tempogan: A temporally coherent,
  volumetric gan for super-resolution fluid flow,'' \emph{ACM Transactions on
  Graphics (TOG)}, vol.~37, no.~4, pp. 1--15, 2018.

\bibitem{werhahn2019multi}
M.~Werhahn, Y.~Xie, M.~Chu, and N.~Thuerey, ``A multi-pass gan for fluid flow
  super-resolution,'' in \emph{ACM Computer Graphics and Interactive
  Techniques}, vol.~2, 2019, pp. 1--21.

\bibitem{bai2020dynamic}
K.~Bai, W.~Li, M.~Desbrun, and X.~Liu, ``Dynamic upsampling of smoke through
  dictionary-based learning,'' \emph{ACM Transactions on Graphics (TOG)},
  vol.~40, no.~1, pp. 1--19, 2020.

\bibitem{sanchez2020learning}
A.~Sanchez-Gonzalez, J.~Godwin, T.~Pfaff, R.~Ying, J.~Leskovec, and
  P.~Battaglia, ``Learning to simulate complex physics with graph networks,''
  in \emph{International Conference on Machine Learning}.\hskip 1em plus 0.5em
  minus 0.4em\relax PMLR, 2020, pp. 8459--8468.

\bibitem{roy2021neural}
B.~Roy, P.~Poulin, and E.~Paquette, ``Neural upflow: A scene flow learning
  approach to increase the apparent resolution of particle-based liquids,''
  \emph{arXiv preprint arXiv:2106.05143}, 2021.

\bibitem{kim2019deep}
B.~Kim, V.~C. Azevedo, N.~Thuerey, T.~Kim, M.~Gross, and B.~Solenthaler, ``Deep
  fluids: A generative network for parameterized fluid simulations,'' in
  \emph{Computer Graphics Forum}, vol.~38, no.~2.\hskip 1em plus 0.5em minus
  0.4em\relax Wiley Online Library, 2019, pp. 59--70.

\bibitem{Chu2021Learning}
M.~Chu, N.~Thuerey, H.-P. Seidel, C.~Theobalt, and R.~Zayer, ``Learning
  meaningful controls for fluids,'' \emph{ACM Transactions on Graphics, (Proc.
  SIGGRAPH)}, vol.~40, no.~4, pp. 100:1--100:13, aug 2021.

\bibitem{wiewel2019latent}
S.~Wiewel, M.~Becher, and N.~Thuerey, ``Latent space physics: Towards learning
  the temporal evolution of fluid flow,'' in \emph{Computer graphics forum},
  vol.~38, no.~2.\hskip 1em plus 0.5em minus 0.4em\relax Wiley Online Library,
  2019, pp. 71--82.

\bibitem{wiewel2020latent}
S.~Wiewel, B.~Kim, V.~C. Azevedo, B.~Solenthaler, and N.~Thuerey, ``Latent
  space subdivision: Stable and controllable time predictions for fluid flow,''
  \emph{arXiv:2003.08723}, 2020.

\bibitem{oyj}
Y.~J. Oh and I.-K. Lee, ``Two-step temporal interpolation network using forward
  advection for efficient smoke simulation,'' in \emph{Computer Graphics
  Forum}, vol.~40, no.~2.\hskip 1em plus 0.5em minus 0.4em\relax Wiley Online
  Library, 2021, pp. 355--365.

\bibitem{ronneberger2015u}
O.~Ronneberger, P.~Fischer, and T.~Brox, ``U-net: Convolutional networks for
  biomedical image segmentation,'' in \emph{International Conference on Medical
  image computing and computer-assisted intervention}.\hskip 1em plus 0.5em
  minus 0.4em\relax Springer, 2015, pp. 234--241.

\bibitem{xiao2018adaptive}
X.~Xiao, C.~Yang, and X.~Yang, ``Adaptive learning-based projection method for
  smoke simulation,'' \emph{Computer Animation and Virtual Worlds}, vol.~29,
  no. 3-4, p. e1837, 2018.

\bibitem{obiols2020cfdnet}
O.~Obiols-Sales, A.~Vishnu, N.~Malaya, and A.~Chandramowliswharan, ``Cfdnet: A
  deep learning-based accelerator for fluid simulations,'' in \emph{Proceedings
  of the 34th ACM International Conference on Supercomputing}, 2020, pp. 1--12.

\bibitem{ragheb1976computational}
M.~Ragheb, ``Computational fluid dynamics,'' \emph{mragheb Website, Available
  Online at https://mragheb. com/NPRE}, vol. 20475, 1976.

\bibitem{9082171}
C.~Li, S.~Qiu, C.~Wang, and H.~Qin, ``Learning physical parameters and detail
  enhancement for gaseous scene design based on data guidance,'' \emph{IEEE
  Transactions on Visualization and Computer Graphics}, vol.~27, no.~10, pp.
  3867--3880, 2021.

\bibitem{10.1145/3478513.3480492}
\BIBentryALTinterwordspacing
K.~Bai, C.~Wang, M.~Desbrun, and X.~Liu, ``Predicting high-resolution
  turbulence details in space and time,'' \emph{ACM Trans. Graph.}, vol.~40,
  no.~6, dec 2021. [Online]. Available:
  \url{https://doi.org/10.1145/3478513.3480492}
\BIBentrySTDinterwordspacing

\bibitem{vae}
D.~P. Kingma and M.~Welling, ``Auto-encoding variational bayes,'' \emph{arXiv
  preprint arXiv:1312.6114}, 2013.

\bibitem{gan}
I.~Goodfellow, J.~Pouget-Abadie, M.~Mirza, B.~Xu, D.~Warde-Farley, S.~Ozair,
  A.~Courville, and Y.~Bengio, ``Generative adversarial nets,'' \emph{Advances
  in neural information processing systems}, vol.~27, 2014.

\bibitem{gulrajani2016pixelvae}
I.~Gulrajani, K.~Kumar, F.~Ahmed, A.~A. Taiga, F.~Visin, D.~Vazquez, and
  A.~Courville, ``Pixelvae: A latent variable model for natural images,''
  \emph{arXiv preprint arXiv:1611.05013}, 2016.

\bibitem{huang2018introvae}
H.~Huang, Z.~Li, R.~He, Z.~Sun, and T.~Tan, ``Introvae: Introspective
  variational autoencoders for photographic image synthesis,'' \emph{arXiv
  preprint arXiv:1807.06358}, 2018.

\bibitem{karras2019style}
T.~Karras, S.~Laine, and T.~Aila, ``A style-based generator architecture for
  generative adversarial networks,'' in \emph{Proceedings of the IEEE/CVF
  Conference on Computer Vision and Pattern Recognition}, 2019, pp. 4401--4410.

\bibitem{choi2020stargan}
Y.~Choi, Y.~Uh, J.~Yoo, and J.-W. Ha, ``Stargan v2: Diverse image synthesis for
  multiple domains,'' in \emph{Proceedings of the IEEE/CVF Conference on
  Computer Vision and Pattern Recognition}, 2020, pp. 8188--8197.

\bibitem{cai2019multi}
L.~Cai, H.~Gao, and S.~Ji, ``Multi-stage variational auto-encoders for
  coarse-to-fine image generation,'' in \emph{Proceedings of the 2019 SIAM
  International Conference on Data Mining}.\hskip 1em plus 0.5em minus
  0.4em\relax SIAM, 2019, pp. 630--638.

\bibitem{brock2018large}
A.~Brock, J.~Donahue, and K.~Simonyan, ``Large scale gan training for high
  fidelity natural image synthesis,'' in \emph{International Conference on
  Learning Representations}, 2018.

\bibitem{jiang2018super}
H.~Jiang, D.~Sun, V.~Jampani, M.-H. Yang, E.~Learned-Miller, and J.~Kautz,
  ``Super slomo: High quality estimation of multiple intermediate frames for
  video interpolation,'' in \emph{Proceedings of the IEEE Conference on
  Computer Vision and Pattern Recognition}, 2018, pp. 9000--9008.

\bibitem{Liu_2017_ICCV}
Z.~Liu, R.~A. Yeh, X.~Tang, Y.~Liu, and A.~Agarwala, ``Video frame synthesis
  using deep voxel flow,'' in \emph{Proceedings of the IEEE International
  Conference on Computer Vision (ICCV)}, Oct 2017.

\bibitem{chen2018lip}
L.~Chen, Z.~Li, R.~K. Maddox, Z.~Duan, and C.~Xu, ``Lip movements generation at
  a glance,'' in \emph{Proceedings of the European Conference on Computer
  Vision (ECCV)}, 2018, pp. 520--535.

\bibitem{Xu_2019_CVPR}
R.~Xu, X.~Li, B.~Zhou, and C.~C. Loy, ``Deep flow-guided video inpainting,'' in
  \emph{Proceedings of the IEEE/CVF Conference on Computer Vision and Pattern
  Recognition (CVPR)}, June 2019.

\bibitem{Wu_2020_CVPR}
Y.~Wu, R.~Gao, J.~Park, and Q.~Chen, ``Future video synthesis with object
  motion prediction,'' in \emph{Proceedings of the IEEE/CVF Conference on
  Computer Vision and Pattern Recognition (CVPR)}, June 2020.

\bibitem{Liang_2017_ICCV}
X.~Liang, L.~Lee, W.~Dai, and E.~P. Xing, ``Dual motion gan for future-flow
  embedded video prediction,'' in \emph{Proceedings of the IEEE International
  Conference on Computer Vision (ICCV)}, Oct 2017.

\bibitem{wang2018video}
T.-C. Wang, M.-Y. Liu, J.-Y. Zhu, G.~Liu, A.~Tao, J.~Kautz, and B.~Catanzaro,
  ``Video-to-video synthesis,'' \emph{arXiv preprint arXiv:1808.06601}, 2018.

\bibitem{lu2021video}
W.~Lu, J.~Cui, Y.~Chang, and L.~Zhang, ``A video prediction method based on
  optical flow estimation and pixel generation,'' \emph{IEEE Access}, vol.~9,
  pp. 100\,395--100\,406, 2021.

\bibitem{ohnishi2018hierarchical}
K.~Ohnishi, S.~Yamamoto, Y.~Ushiku, and T.~Harada, ``Hierarchical video
  generation from orthogonal information: Optical flow and texture,'' in
  \emph{Thirty-Second AAAI Conference on Artificial Intelligence}, 2018.

\bibitem{Li_2018_ECCV}
Y.~Li, C.~Fang, J.~Yang, Z.~Wang, X.~Lu, and M.-H. Yang, ``Flow-grounded
  spatial-temporal video prediction from still images,'' in \emph{Proceedings
  of the European Conference on Computer Vision (ECCV)}, September 2018.

\bibitem{pan2019video}
J.~Pan, C.~Wang, X.~Jia, J.~Shao, L.~Sheng, J.~Yan, and X.~Wang, ``Video
  generation from single semantic label map,'' in \emph{Proceedings of the
  IEEE/CVF Conference on Computer Vision and Pattern Recognition}, 2019, pp.
  3733--3742.

\bibitem{wang2020vidsr}
L.~Wang, Y.~Guo, L.~Liu, Z.~Lin, X.~Deng, and W.~An, ``Deep video
  super-resolution using hr optical flow estimation,'' \emph{IEEE Transactions
  on Image Processing}, vol.~29, pp. 4323--4336, 2020.

\bibitem{wei2020learning}
Y.~Wei, M.~Liu, H.~Wang, R.~Zhu, G.~Hu, and W.~Zuo, ``Learning flow-based
  feature warping for face frontalization with illumination inconsistent
  supervision,'' in \emph{European Conference on Computer Vision}.\hskip 1em
  plus 0.5em minus 0.4em\relax Springer, 2020, pp. 558--574.

\bibitem{vgg}
K.~Simonyan and A.~Zisserman, ``Very deep convolutional networks for
  large-scale image recognition,'' \emph{arXiv preprint arXiv:1409.1556}, 2014.

\bibitem{johnson2016perceptual}
J.~Johnson, A.~Alahi, and L.~Fei-Fei, ``Perceptual losses for real-time style
  transfer and super-resolution,'' in \emph{European conference on computer
  vision}.\hskip 1em plus 0.5em minus 0.4em\relax Springer, 2016, pp. 694--711.

\bibitem{deng2009imagenet}
J.~Deng, W.~Dong, R.~Socher, L.-J. Li, K.~Li, and L.~Fei-Fei, ``Imagenet: A
  large-scale hierarchical image database,'' in \emph{2009 IEEE conference on
  computer vision and pattern recognition}.\hskip 1em plus 0.5em minus
  0.4em\relax Ieee, 2009, pp. 248--255.

\bibitem{chang2019free}
Y.-L. Chang, Z.~Y. Liu, K.-Y. Lee, and W.~Hsu, ``Free-form video inpainting
  with 3d gated convolution and temporal patchgan,'' in \emph{Proceedings of
  the IEEE/CVF International Conference on Computer Vision}, 2019, pp.
  9066--9075.

\bibitem{vougioukas2020realistic}
K.~Vougioukas, S.~Petridis, and M.~Pantic, ``Realistic speech-driven facial
  animation with gans,'' \emph{International Journal of Computer Vision}, vol.
  128, no.~5, pp. 1398--1413, 2020.

\bibitem{pytorch}
A.~Paszke, S.~Gross, F.~Massa, A.~Lerer, J.~Bradbury, G.~Chanan, T.~Killeen,
  Z.~Lin, N.~Gimelshein, L.~Antiga, A.~Desmaison, A.~Kopf, E.~Yang, Z.~DeVito,
  M.~Raison, A.~Tejani, S.~Chilamkurthy, B.~Steiner, L.~Fang, J.~Bai, and
  S.~Chintala, ``Pytorch: An imperative style, high-performance deep learning
  library,'' in \emph{Advances in Neural Information Processing Systems 32},
  H.~Wallach, H.~Larochelle, A.~Beygelzimer, F.~d\textquotesingle
  Alch\'{e}-Buc, E.~Fox, and R.~Garnett, Eds.\hskip 1em plus 0.5em minus
  0.4em\relax Curran Associates, Inc., 2019, pp. 8024--8035.

\bibitem{kingma2014adam}
D.~P. Kingma and J.~Ba, ``Adam: A method for stochastic optimization,''
  \emph{arXiv preprint arXiv:1412.6980}, 2014.

\bibitem{mantaflow}
\BIBentryALTinterwordspacing
N.~Thuerey and T.~Pfaff, ``{MantaFlow},'' 2018. [Online]. Available:
  \url{http://mantaflow.com}
\BIBentrySTDinterwordspacing

\bibitem{chu2020tecoGAN}
M.~Chu, Y.~Xie, J.~Mayer, L.~Leal-Taixe, and N.~Thuerey, ``{Learning Temporal
  Coherence via Self-Supervision for GAN-based Video Generation (TecoGAN)},''
  \emph{ACM Transactions on Graphics}, vol.~39, no.~4, 2020.

\bibitem{10.5555/2422356.2422373}
M.~Lentine, M.~Cong, S.~Patkar, and R.~Fedkiw, ``Simulating free surface flow
  with very large time steps,'' in \emph{Proceedings of the ACM
  SIGGRAPH/Eurographics Symposium on Computer Animation}, ser. SCA '12.\hskip
  1em plus 0.5em minus 0.4em\relax Goslar, DEU: Eurographics Association, 2012,
  p. 107–116.

\bibitem{10.1145/2019406.2019419}
\BIBentryALTinterwordspacing
M.~Lentine, M.~Aanjaneya, and R.~Fedkiw, ``Mass and momentum conservation for
  fluid simulation,'' in \emph{Proceedings of the 2011 ACM
  SIGGRAPH/Eurographics Symposium on Computer Animation}, ser. SCA '11.\hskip
  1em plus 0.5em minus 0.4em\relax New York, NY, USA: Association for Computing
  Machinery, 2011, p. 91–100. [Online]. Available:
  \url{https://doi.org/10.1145/2019406.2019419}
\BIBentrySTDinterwordspacing

\end{thebibliography}

%




\end{document}